\documentclass[12pt]{article}

\usepackage[margin=1in]{geometry}  
\usepackage{graphicx}              
\usepackage{amsmath}               
\usepackage{amsfonts}              
\usepackage{amsthm}                

\newcommand{\Ai}{\mbox{Ai}}
\newcommand{\tr}{\mbox{tr}}

\begin{document}

\nocite{*}

\title{Path Integral Monte-Carlo Calculations for Relativistic Oscillator}

\author{Alexandr Ivanov, Oleg Pavlovsky \\
Faculty of Physics, Lomonosov Moscow State University}

\maketitle

\begin{abstract}
The problem of Relativistic Oscillator has been studied in the framework of Path Integral Monte-Carlo(PIMC) approach. Ultra-relativistic and non-relativistic limits have been discussed. We show that PIMC method can be effectively used for investigation of relativistic systems.
\end{abstract}

\section{Introduction}

Path Integral Monte-Carlo method \cite{Condensed helium} is one of the
most popular {\it ab initio} numerical approach of the investigation of
quantum systems. Especially this method becomes useful for
modelling properties of quantum problems of many bodies in
case the Schrodinger Equations become difficult to study
but the number of quantum degrees of freedom is still not so large
for using Quantum Statistics.

This work is devoted to the generalization of Path Integral Monte
Carlo in case of the relativistic systems. There are many
physical problems connected with simulations of relativistic quantum
mechanical systems. Relativistic corrections play very essential
role in physics of the atomic systems with heavy elements
due to the strong interaction potentials. One may find the problems with
simulations of relativistic quantum systems in the
nuclear physics, physics of hadron structure and quark-gluon
plasma, in relativistic astrophysics. Recently an another
interesting application of the relativistic quantum mechanics has arisen. These are so called (Pseudo) Relativistic Condensed
Matter systems and one of the most famious examples of such systems
is Graphene \cite{Novoselov:04:1}.

The remarkable property of Graphene is that it's effective
charge excitations have very small mass (in case of the
idealistic graphene without defects and boundary effects the mass
of excitations is equal to zero). On the other hand, the
interaction between excitations is very strong. It means that the
excitations on the graphene sheet can be treated as some strongly
interacted two-dimensional relativistic gas with instantaneous
interaction.

The correct formulation of the many-body relativistic
quantum-mechanical  problem has some well-known difficulties.
The kinetic and potential part of the Hamiltonian must be
invariant at Lorentz transformation. The kinetic part of
Hamiltonian can be formulated in Lorentz-invariant form relatively
easy but relativistic formulation of interaction requires the
quantum field theory approach in general case. If we want to work
in the framework of the quantum mechanics approach we are to use
some additional assumptions. In this work we study just kinetic part of Hamiltonian. The interaction part
can not be studied in general case. For the first approximation let us
consider the relativistic quantum systems with instantaneous
interaction between particles. This approximation works very well
in case of Relativistic Quantum Chemistry and for
investigation of the properties (Pseudo) Relativistic Condensed
Matter systems like graphene where the correction from
relativistic nature of the interaction is fortunately very small in
comparison with the correction that comes from relativistic nature of
the particles. The nuclear and high-energy systems require
some special considerations and these tasks will be the key point of
our future works.

Relativistic generalization of the Path Integral approach for
quantum mechanical systems has a long history \cite{Fiziev},
\cite{Redmount}. Today this approach is becoming more and more popular
and find its application in high-energy physics \cite{Filinov1},
\cite{Filinov2}. Now we briefly discuss the main statements of Path Integral
formalism for relativistic quantum-mechanical system.

\section{Oscillator}
Simple harmonic oscillator has a hamiltonian function
\begin{equation} \label{SHO}
H = m + \frac{p ^ 2}{2 m} + \frac{1}{2} m \omega ^ 2 q ^ 2,
\end{equation}
where $V (q) = \frac{1}{2} m \omega ^ 2 q ^ 2$ - potential energy, $T (p) = \frac{p ^ 2}{2 m}$ - kinetic energy and $m$ - rest mass. The expression $\frac{p ^ 2}{2 m}$ is kinetic energy in non-relativistic case. Generalization of kinetic energy is $T (p) = \sqrt{p ^ 2 + m ^ 2}$. So the generalization of simple harmonic oscillator is the following
\begin{equation} \label{Rel}
H = \sqrt{p ^ 2 + m ^ 2} + \frac{1}{2} m \omega ^ 2 q ^ 2.
\end{equation}

In this work we want to calculate the system with hamiltonian function $(\ref{Rel})$ by Path Integral Monte-Carlo(PIMC) Metropolis algorithm, but at first, this method should be generalized for relativistic case. To compare the results we have used two limits of this hamiltonian function, in which we have analytical expressions for observables:

\begin{equation}\label{Nonrel}
m ^ 2 \gg p ^ 2 \text{ (Non-relativistic limit)}
\end{equation}
\begin{equation}\label{Ultrarel}
m ^ 2 \ll p ^ 2 \text{ (Ultra-relativistic limit)}
\end{equation}

There are only two dimensional parameters $m$ and $\omega$, as we will see later, conditions $(\ref{Nonrel})$ and $(\ref{Ultrarel})$ can be overwrite in terms of $m$ and $\omega$.
\subsection{Non-relativistic limit}
In this section the limit $m ^ 2 \gg p ^ 2$ is considered. We should understand it in terms of average values, it means that $m ^ 2 \gg \langle p ^ 2 \rangle$.
Non-relativistic limit of the kinetic energy from ($\ref{Rel}$) is the following
$$
T(p) = \sqrt{p ^ 2 + m ^ 2} = m (1 + \frac{p ^ 2}{2 m ^ 2} + O \Bigl ( \Bigl ( \frac{p}{m} \Bigr ) ^ 4 \Bigr ) ) \approx m + \frac{p ^ 2}{2 m}
$$
and we obtain simple harmonic oscillator. 
\begin{equation}\label{hamiltonnonrel}
H = m + \frac{p ^ 2}{2 m} + \frac{1}{2} m \omega ^ 2 q ^ 2.
\end{equation}
The energy and probability density of ground state is well known for this hamiltonian function
\begin{equation}\label{energynonrel}
E _ 0 = m + \frac{\omega}{2},
\end{equation}
\begin{equation}\label{densitynonrel}
\rho (q) = |\psi (q)| ^ 2 = \Bigl ( \frac{m \omega}{\pi} \Bigr ) ^ \frac{1}{2} \exp \Bigl ( -m \omega q ^ 2 \Bigr ), 
\end{equation}
where $\psi (q)$ is a wave function of this ground state.
The virial theorem gives us relation between kinetic and potential energy
$$
\Bigl \langle \frac{p ^ 2}{2 m} \Bigr \rangle = \langle \frac{1}{2} m \omega ^ 2 q ^ 2 \rangle \text{ or } \langle T(p) - m \rangle = \langle V (q) \rangle.
$$
Using $(\ref{hamiltonnonrel})$ we obtain
\begin{equation}\label{virialnonrel}
\langle T (p) - m \rangle = \frac{1}{2} (E _ 0 - m) = \langle V (q) \rangle = \frac{\omega}{4}.
\end{equation}
Correlation function $\langle q(t) q(t + s) \rangle$ for this system is the following
\begin{equation}\label{correlationnonrel}
\langle q(t) q(t + s) \rangle = \frac{1}{2 m \omega} e ^ {- w |s|}.
\end{equation}
We have a list of observables $(\ref{energynonrel})$, $(\ref{densitynonrel})$, $(\ref{correlationnonrel})$ to compare results with numerical calculations by PIMC.

We can formulate condition $(\ref{Nonrel})$ in terms of $m$ and $\omega$, that is necessary to distinguish different limits. Using $(\ref{Nonrel})$, $(\ref{virialnonrel})$ we can obtain $\langle p ^ 2 \rangle \sim m \omega$ and $m \gg \omega$. It means that we consider heavy particles and soft potential in this case. 

\subsection{Ultra-relativistic limit}
Our next step is a consideration of hamiltonian function $(\ref{Rel})$ in the limit $(\ref{Ultrarel})$.
The kinetic energy in this case is the following
$$
T (p) = \sqrt{p ^ 2 + m ^ 2} = |p| \Bigl ( 1 + \frac{m ^ 2}{2 p ^ 2} + O \Bigl ( \Bigl ( \frac{m}{p} \Bigr ) ^ 4 \Bigr ) \Bigr ) \approx |p|.
$$
That's enough to take only zero order to describe the behavior of the system. We have
\begin{equation}\label{hamiltinianulrel}
H = |p| + \frac{1}{2} m \omega ^ 2 q ^ 2.
\end{equation}
We can solve Shroedinger equation for this hamiltonian function in momentum representation and find energy of ground state and corresponding density of probability
\begin{equation}\label{energyulrel}
E _ 0 = \lambda _ 0 (m \omega ^ 2) ^ {1 / 3},
\end{equation}
where $\lambda _ 0 = 0.808617\dots$, 
\begin{equation}\label{densityulrel}
\rho (q) = \frac{\int \int \frac{dp dk}{(2 \pi) ^ 2} \Ai \Bigl ( (\frac{2}{m \omega ^ 2}) ^ {1 / 3} (|p| - \lambda _ 0) \Bigr )\Ai \Bigl ( (\frac{2}{m \omega ^ 2}) ^ {1 / 3} (|k| - \lambda _ 0) \Bigr )  e ^ {-i (p - k) q}}{\int \frac{dp}{2 \pi} \Ai ^ 2 \Bigl ( (\frac{2}{m \omega ^ 2}) ^ {1 / 3} (|p| - \lambda _ 0) \Bigr )}, 
\end{equation}
where $\Ai(x)$ is the Airy function.
The virial theorem for this hamiltonian function gives us the relation between kinetic and potential energy
$$
\langle T (p) \rangle = 2 \langle V (q) \rangle,
$$ 
so we can obtain results for kinetic and potential energy
\begin{equation}\label{kineticulrel}
\langle T (p) \rangle = \frac{2 \lambda _ 0}{3} (m \omega ^ 2) ^ {1 / 3},
\end{equation}
\begin{equation}\label{potentialulrel}
\langle V (q) \rangle = \frac{\lambda _ 0}{3} (m \omega ^ 2) ^ {1 / 3},
\end{equation}
\begin{equation}\label{x2ulrel}
\langle q ^ 2 \rangle = \frac{2 \lambda _ 0}{3 (m \omega ^ 2) ^ {2 / 3}}.
\end{equation}
Using the virial theorem, we have $\langle |p| \rangle \sim (m \omega ^ 2) ^ {1 / 3}$. Let's suggest $\langle |p| \rangle ^ 2 \sim \langle p ^ 2 \rangle$, than we can obtain  ratio
$$
\frac{m ^ 2}{\langle p ^ 2 \rangle} \sim \Bigl ( \frac{m}{\omega} \Bigr ) ^ {4 / 3} \ll 1.
$$
For ultra-relativistic case we have the opposite expression for mass and frequency
$$
\omega \gg m.
$$
We have obtained one more list of observables for comparison with PIMC calculations.

\section{Density matrix for Monte-Carlo calculations}
In this section we consider quantum mechanics system at finite temperature. One can find the full consideration  of this question in $\cite{Condensed helium}$. Average value of some operator $A$
$$
    \langle A \rangle = \frac{\tr (A e ^ {-\beta H})}{\tr (e ^ {-\beta H})},
$$
where operator $e ^ {-\beta H}$ is the density matrix and $\beta = 1 / \theta$ is inverse temperature of the system to be considered. For PIMC method we should consider zero temperature limit. Matrix element of density matrix in coordinate representation is
$$
    \rho (q, q '; \beta) = \langle q | e ^ {-\beta H} | q ' \rangle.
$$
Finally, we have for operator $A$
$$
    \langle A \rangle = \frac{\int dq dq ' \rho (q, q' ; \beta) \langle q | A | q ' \rangle}{\int dq \rho (q, q ; \beta)}.
$$
For density matrix operator we can write expression
$$
    e ^ {-(\beta _ 1 + \beta _ 2) H} = e ^ {-\beta _ 1 H} e ^ {-\beta _ 2 H}
$$
and the same in coordinate representation
$$
    \rho (q _ 1, q _ 3; \beta _ 1 + \beta _ 2) = \int d q _ 2 \rho (q _ 1, q _ 2; \beta _ 1) \rho (q _ 2, q _ 3; \beta _ 2).
$$
Applied this property $N _ t$ times
$$
    e ^ {- \beta H} = (e ^ {- \tau H}) ^ {N _ t},
$$
where $\tau = \beta / N _ t$.  And in coordinate representation
$$
    \rho (q _ 0, q _ N; \beta) = \int \dots \int dq _ 1 dq _ 2 \dots dq _ {N - 1} \rho(q _ 0, q _ 1; \tau) \rho (q _ 1, q _ 2; \tau) \dots \rho(q _ {N - 1}, q _ N; \tau).
$$
Operator of kinetic energy is diagonal in momentum representation and operator of potential energy is diagonal in coordinate representation. We can separate kinetic and potential energy if $\tau$ is small, so
$$
    e ^ {-\tau (T + V) + \frac{\tau ^ 2}{2} [T, V]} = e ^ {-\tau T} e ^ {-\tau V}.
$$
And if $\tau \to 0$
$$
    e ^ {-\tau (T + V)} \approx e ^ {-\tau T} e ^ {-\tau V}.
$$
$$
    \rho (q _ 0, q _ 2; \tau) \approx \int dq _ 1 \langle q _ 0 | e ^ {-\tau T} | q _ 1 \rangle \langle q _ 1 | e ^ {-\tau V} | q _ 2 \rangle.
$$
Potential energy is diagonal in position representation, so
$$
    \langle q _ 1 | e ^ {-\tau V} | q _ 2 \rangle = e ^ {-\tau V (q _ 1)} \delta (q _ 2 - q _ 1)    .
$$
Using the expression $\int | p \rangle \frac{dp}{2 \pi} \langle p | = 1$ we obtain
$$
    \langle q _ 0 | e ^ {-\tau T} | q _ 1 \rangle = \int dp dp ' \delta (p - p') \langle q _ 0 | p \rangle \langle p ' | q _ 1 \rangle e ^ {-T (p) \tau}.
$$
Taking into account $\langle q | p \rangle = e ^{-i q p}$, we obtain
$$
    \langle q _ 0 | e ^ {-\tau T} | q _ 1 \rangle = \int \frac{dp}{2 \pi} e ^ {-T (p) \tau -i p (q _ 0 - q _ 1)}.
$$
So to build path integral we must take this integral over momentum.
Let's consider this integral in general case or, in other words, with general relativistic kinetic energy. We can calculate integral over momentum with this $T(p)$
$$
\langle q _ 0 | e ^ {-\tau T} | q _ 1 \rangle = \frac{m \tau}{\pi \sqrt{\tau ^ 2 + (q _ 1 - q _ 0) ^ 2}} K _ 1 (m \sqrt{\tau ^ 2 + (q _ 1 - q _ 0) ^ 2}) = 
$$
$$
= \frac{m}{\pi \sqrt{1 + \Bigl ( \frac{q _ 1 - q _ 0}{\tau} \Bigr ) ^ 2}} K _ 1 (m \tau \sqrt{1 + \Bigl ( \frac{q _ 1 - q _ 0}{\tau} \Bigr ) ^ 2}).
$$
where $K _ 1 (*)$ is modified Bessel function of the first order. The general expression for matrix element is the following
\begin{equation}\label{matrixelem}
\rho (q '', q '; \tau) = \langle q '' | e ^ {-\tau (T(p) + V(q))} | q ' \rangle = \frac{m}{\pi \sqrt{1 + \Bigl ( \frac{q '' - q '}{\tau} \Bigr ) ^ 2}} K _ 1 \Bigl [ m \tau \sqrt{1 + \Bigl ( \frac{q '' - q '}{\tau} \Bigr ) ^ 2} \Bigr ] e ^ {-\tau V(q ')}.
\end{equation}

\section{Path Integral Monte-Carlo Metropolis Algorithm}
For Path Integral Monte-Carlo Metropolis algorithm we should know a part of the density matrix which corresponds to fixed point $q _ i$. Discussion about this method and all proofs one can find in $\cite{Creutz}$. Using ($\ref{matrixelem}$) we can write
$$
\pi (q _ i) = \frac{m ^ 2 K _ 1 \Bigl [ m \tau \sqrt{1 + \Bigl ( \frac{q _ i - q _ {i - 1}}{\tau} \Bigr ) ^ 2} \Bigr ] K _ 1 \Bigl [ m \tau \sqrt{1 + \Bigl ( \frac{q _ {i + 1} - q _ i}{\tau} \Bigr ) ^ 2} \Bigr ]}{\pi ^ 2 \sqrt{1 + \Bigl ( \frac{q _ i - q _ {i - 1}}{\tau} \Bigr ) ^ 2} \sqrt{1 + \Bigl ( \frac{q _ {i + 1} - q _ i}{\tau} \Bigr ) ^ 2}} e ^ {-\tau V(q _ i)}.
$$
So to calculate path integral we construct Markov chain which has equilibrium state for fixed $q _ i$ proportional to $\pi (q _ i)$. Transition probability for this Markov chain satisfies equation
$$
\int d q _ i \pi (q _ i) P(q _ i \to q _ i ') = \pi (q _ i ').
$$
We want to obtain $\pi (q)$ as limit of Markov chain, so we require the detailed balance
$$
\pi (q _ i) P (q _ i \to q _ i ') = \pi (q _ i ') P (q _ i ' \to q _ i).
$$
In Metropolis algorithm we can split transition probability
$$
P(q _ i \to q _ i ') = T (q _ i \to q _ i ') A (q _ i \to q _ i '),
$$
where $T (q _ i \to q _ i ')$ is sampling distribution and $A (q _ i \to q _ i ')$ is acceptance probability
$$
A (q _ i \to q_ i ') = \text{min} \Bigl [1, \frac{T(q _ i ' \to q _ i) \pi (q _ i ')}{T(q _ i \to q _ i ') \pi (q _ i)} \Bigr ].
$$
If we know density matrix in position representation, we can calculate expressions for average of any observables. Using ($\ref{matrixelem}$) we can obtain the average of kinetic energy in general relativistic case (see $\cite{Rothe}$), 
\begin{equation}\label{meankineticgenrel}
\langle T(p) \rangle = \Bigl \langle \frac{m \tau}{\sqrt{\tau ^ 2 + (\Delta q) ^ 2}} \frac{K _ 0 (m \sqrt{\tau ^ 2 + (\Delta q) ^ 2})}{K _ 1 (m \sqrt{\tau ^ 2 + (\Delta q) ^ 2})} + \frac{\tau ^ 2 - (\Delta q) ^ 2}{\tau (\tau ^ 2 + (\Delta q) ^ 2)} \Bigr \rangle.
\end{equation}
Average of potential energy
\begin{equation}\label{meanpotentialgenrel}
\langle V(q) \rangle = \Bigl \langle \frac{1}{2} m \omega ^ 2 q _ i ^ 2 \Bigr \rangle.
\end{equation}
Full energy of ground state
$$
\langle E(p, q) \rangle = \langle T(p) + V(q) \rangle.
$$
Probability density
\begin{equation}\label{probdens}
\rho (q) = |\psi (q)| ^ 2 = \frac{1}{N \Delta q} \sum _ {\text{all paths}} \theta(\Delta q - |q - q _ i|),
\end{equation}
where $N$ - all simulation points. Correlation function
\begin{equation}\label{corrfunction}
\langle q (t) q (t + n \tau) \rangle = \langle q _ i q _ {i + n} \rangle.
\end{equation}
We have expressions for generating Markov chain and for calculating observables. Now we can compare results obtained by Monte-Carlo simulations and theretical predictions.

\subsection{Non-relativistic limit}
In this part we compare resuts obtained by PIMC program for $m \gg \omega$ case with theoretical predictions for harmonic oscillator. It is important to find proper values for the program because the quality of the results depends on them. Taking into account features of PIMC Metropolis algotithm it is necessary to reach the limit of Markov chain and desired level of errors to compare results with analytical expressions. Designations to be used in this paper are the following: $N _ p$ - count of paths to be generated, $N _ s$ - count of sweeps for paths generation, $N _ t$ - count of time slices, $\tau$ - time step, $N$ - count of attempts to change any $q _ i$. Following data obtained at this parameters of program $N _ p = 1000, N _ s = 5000, N _ t = 100, \tau = 0.1, N = 10.$

\begin{figure}[h]
\noindent\centering{
\includegraphics[width=80mm]{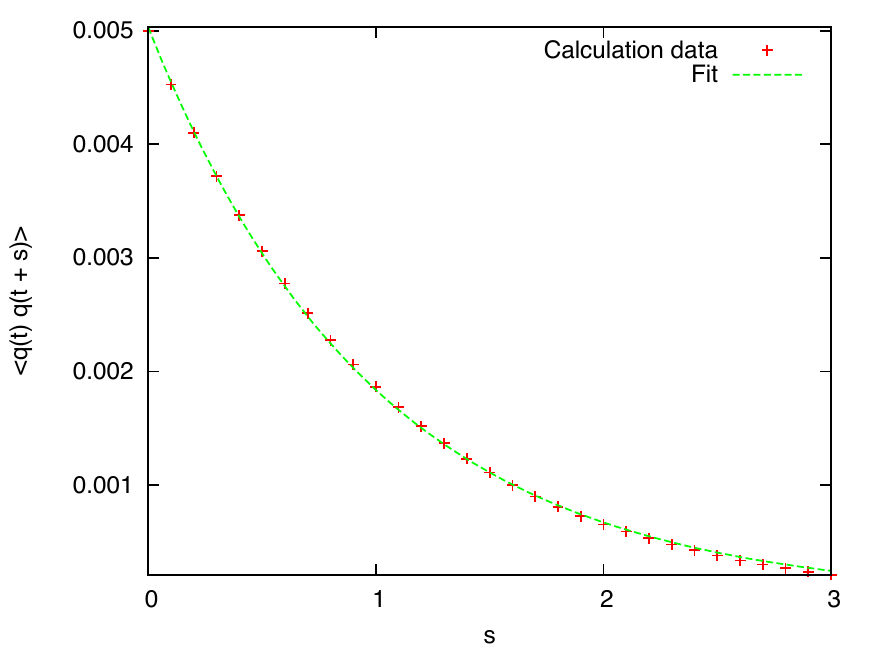}
\caption{Correlation function for $m = 100$, $\omega = 1$.}}
\label{figCurves}
\end{figure}

For correlation function we have good agreement with theoretical expression ($\ref{correlationnonrel}$) (see Fig. 1). Fit with function is the following
$$
\langle q (t) q (t + s) \rangle = a \exp{(-b |s|)}
$$
where obtained values are $a = 1.006 \pm 0.004$ and $b = (503.0 \pm 1.4) \times 10 ^ {-5}$. We know these values for harmonic oscillator $a = m \omega = 1$ and $b = 1 / (2 m \omega) = 500 \times 10 ^ {-5}$.

Let's consider the comparison of results for $\langle q ^ 2 (t) \rangle$ (see Fig. 2).
\begin{figure}[h]
\noindent\centering{
\includegraphics[width=80mm]{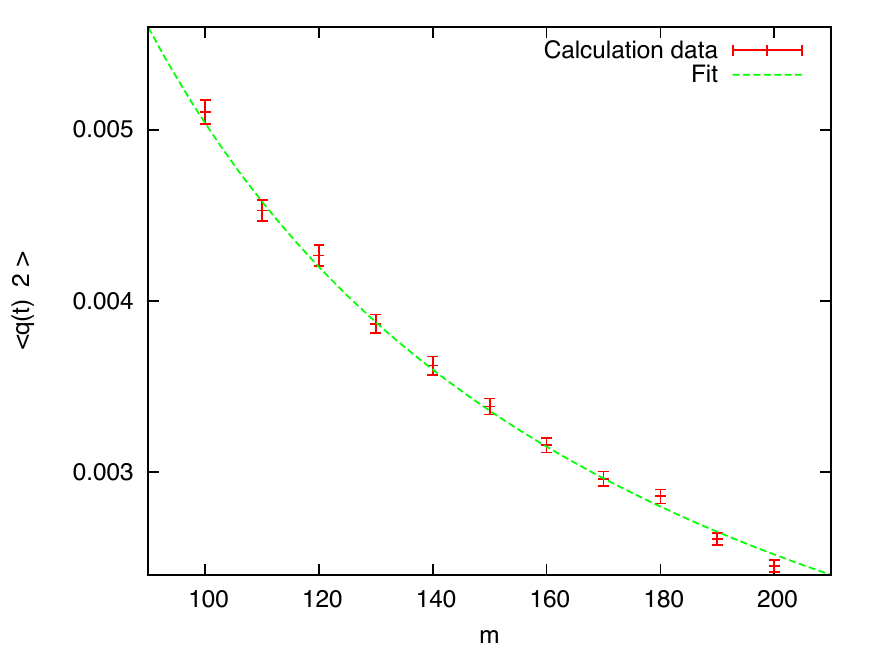}
\caption{Dependence of $\langle q ^ 2 (t) \rangle$ on mass for $\omega = 1$.}}
\label{figCurves}
\end{figure}
\bibliographystyle{plain}
Fit with function is the following
$$
\langle q ^ 2 (t) \rangle = \frac{a}{m}
$$
where for $a$ we obtained $a = 0.5040 \pm 0.0022$, and for harmonic oscillator $a = 1 / (2 \omega) = 0.5$.

Let's consider results of PIMC program for energy, we have analytical expressions ($\ref{virialnonrel}$) (see Fig. 3).
\begin{figure}[h]
\noindent\centering{
\includegraphics[width=80mm]{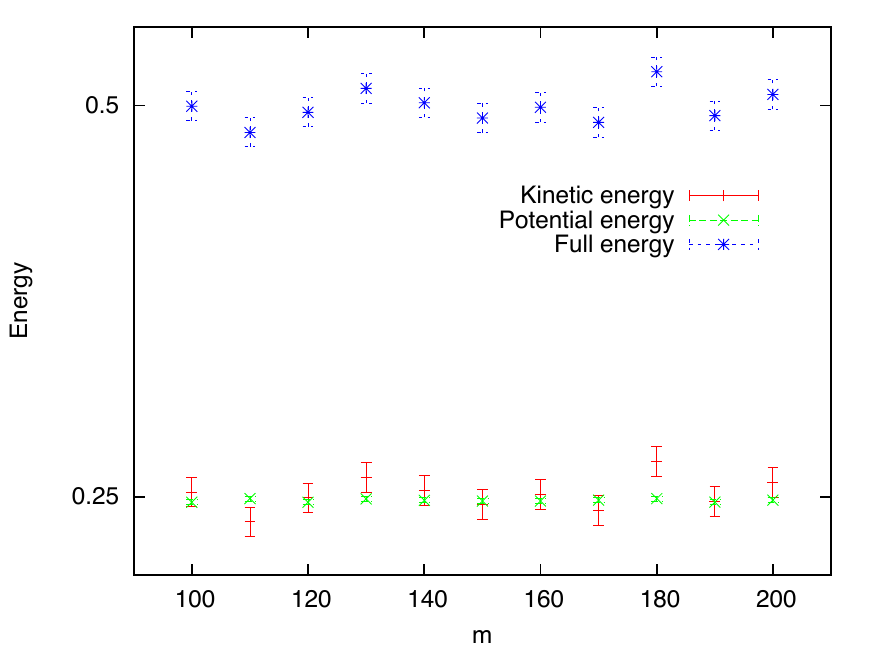}
\caption{Dependence of Energy on mass for $\omega = 1$.}}
\label{figCurves}
\end{figure}
\bibliographystyle{plain}

Fit with constant for this data
$$
\langle T (p) \rangle = a
$$
$$
\langle V (q) \rangle = b
$$
gives us $a = 0.251 \pm 0.003$, $b = 0.2476 \pm 0.0003$, where theoretical predictions are $a = b = \omega / 4 = 0.25$.

We can compare density of probability with ($\ref{densitynonrel}$) (see Fig. 4).
\begin{figure}[h]
\noindent\centering{
\includegraphics[width=80mm]{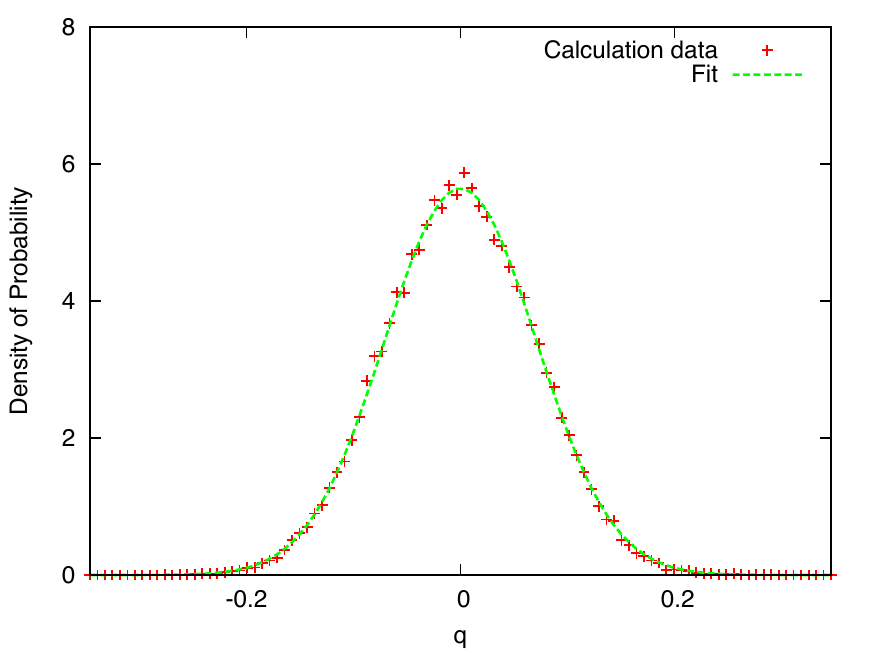}
\caption{Dependence of Density of Probability $|\psi (q)| ^ 2$ on q for $m = 100$, $\omega = 1$.}}
\label{figCurves}
\end{figure}
\bibliographystyle{plain}

Fit with Gaussian function
$$
\rho (q) = a \exp{(-b q ^ 2)}
$$
gives $a = 100.1 \pm 0.7$ and $b = 99.5 \pm 0.8$. We know that in non-relativistic limit $a = b = m \omega = 100$.

As we can see, we have an agreement with simple harmonic oscillator in the limit of $m \gg \omega$ or non-relativistic limit. Somewhere we have agreement with two or more standard deviations, it means that we should increase ratio of $m$ and $\omega$, but computer time of calculations for this case is longer.

\subsection{Ultra-relativistic limit}
In this section we compare results of PIMC Metropolis with $p ^ 2 \gg m ^ 2$ or $\omega \gg m$ limit. Parameters of program to be used for calculations are the following $N _ p = 100, N _ s = 5000, N _ t = 1000, \tau = 0.01, N = 10$.

For correlation function we have (see Fig. 5).
\begin{figure}[h]
\noindent\centering{
\includegraphics[width=80mm]{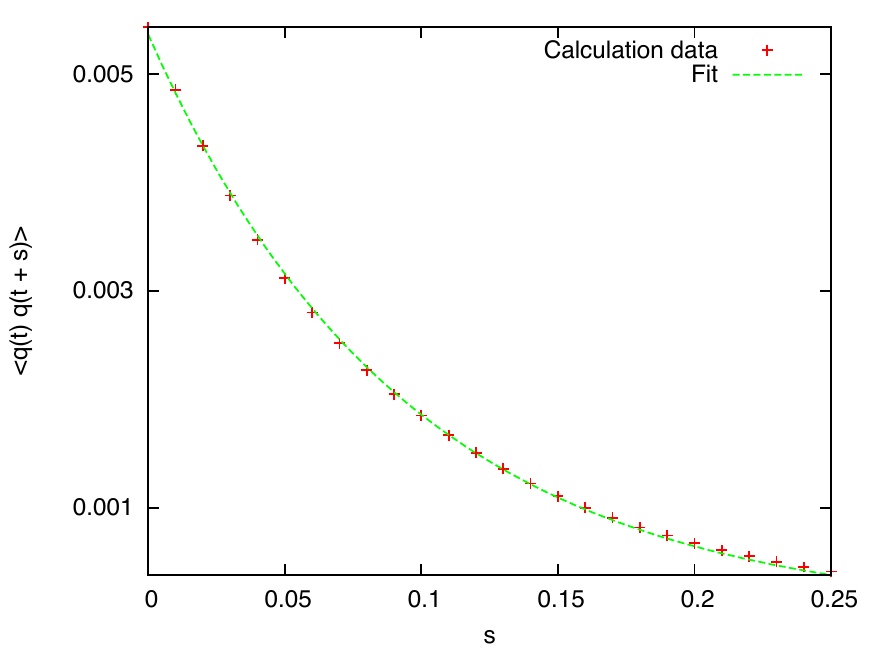}
\caption{Dependence of Correlation function $\langle q (t) q (t + s) \rangle$ on $|s|$ for $m = 0.1$, $\omega = 100$.}}
\label{figCurves}
\end{figure}
\bibliographystyle{plain}

Fit with exponential function
$$
\langle q (t) q (t + s) \rangle = a \exp{(-b |s|)}
$$
gives us $a = (536.5 \pm 1.8) \times 10 ^ {-5}$ and $b = 10.60 \pm 0.05$. Although we did not find analytical expression for correlation function, there are a lot of arguments that should be the following
$$
\langle q (t) q (t + s) \rangle = q ^ 2 (0) e ^ {-(\lambda _ 1 - \lambda _ 0) (m \omega ^ 2) ^ {1/3}}
$$
where $\lambda _ 1 (m \omega ^ 2) ^ {1 / 3}$ is the next energy level of the ultra-relativistic oscillator, and $\lambda _ 1 = 2.338 \dots$. Theoretical values are $a = 0.539$ and energy gap $b = 10.47$.  We have an agreement with theoretical expression for correlation function, so we can continue comparison of results.

Let's consider average of $q ^ 2 (t)$. We have an analytical expression
$$
\langle q ^ 2 (t) \rangle = \frac{2 \lambda _ 0}{3 (m \omega ^ 2) ^ {2 / 3}}
$$
and Monte-Carlo calculations give us following result (see Fig. 6)
\begin{figure}[h]
\noindent\centering{
\includegraphics[width=80mm]{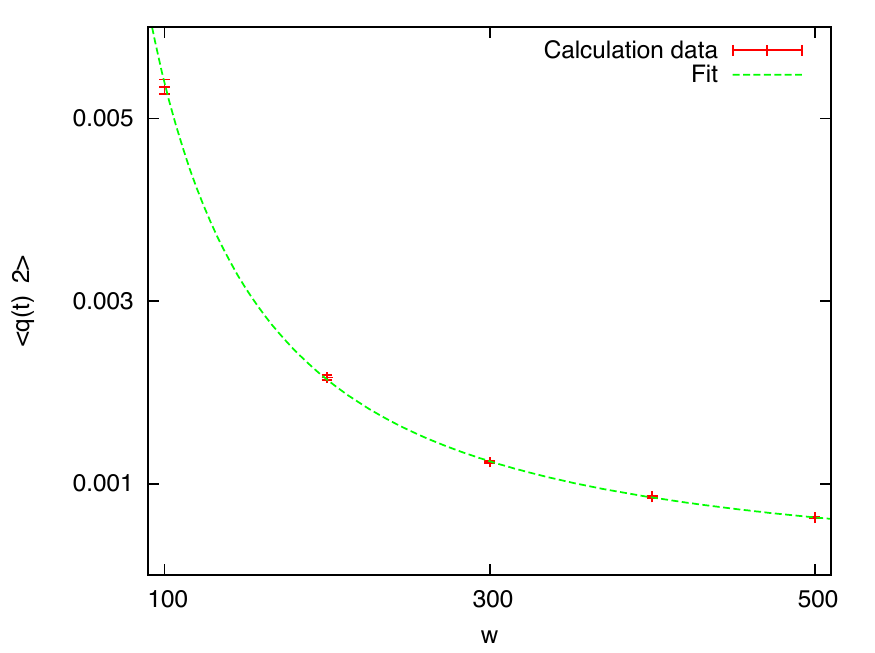}
\caption{Dependence of $\langle q ^ 2 (t) \rangle$ on $\omega$ for $m = 0.1$.}}
\label{figCurves}
\end{figure}

Fit with function
$$
\langle q ^ 2 (t) \rangle = \frac{2 a}{3 (0.1 x ^ 2) ^ {2 / 3}}
$$
gives us $a = 0.8088 \pm 0.0027$, and theoretical prediction is $a = \lambda _ 0 = 0.8086 \dots$

Let us consider the calculations of energy for ultra-relativistic oscillator and compare them with analytical expressions ($\ref{kineticulrel}$), ($\ref{potentialulrel}$) (see Fig. 7).
\begin{figure}[h]
\noindent\centering{
\includegraphics[width=80mm]{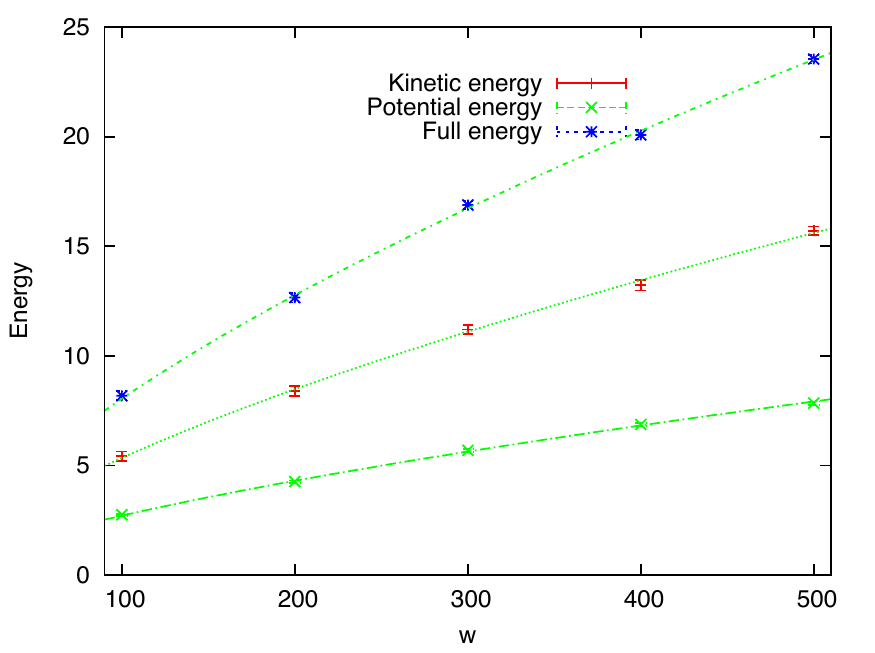}
\caption{Dependence of Energy on $\omega$ for $m = 0.1$.}}
\label{figCurves}
\end{figure}

Fit with functions
$$
\langle T (p) \rangle = \frac{2}{3} a (m \omega ^ 2) ^ {1 / 3},
$$
$$
\langle V (q) \rangle = \frac{1}{3} b (m \omega ^ 2) ^ {1 / 3},
$$
$$
\langle E (p, q) \rangle = c (m \omega ^ 2) ^ {1 / 3}
$$
gives the following values $a = 0.801 \pm 0.004$, $b = 0.812 \pm 0.004$ and $c = 0.804 \pm 0.003$. Theoretical values are $a = b = c = \lambda _ 0 = 0.8086 \dots$.

At the end of this section we compare Monte-Carlo calculations of density of probability with theoretical prediction ($\ref{densityulrel}$) (see Fig. 8).
\begin{figure}[h]
\noindent\centering{
\includegraphics[width=80mm]{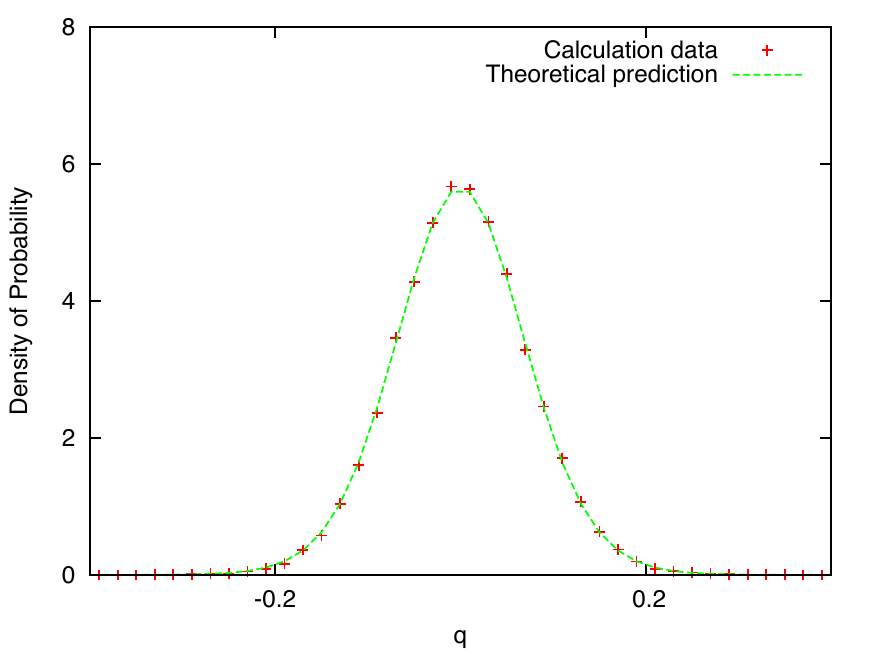}
\caption{Dependence of Density of Probability on $q$ for $m = 0.1$ and $\omega = 100$.}}
\label{figCurves}
\end{figure}

As a result, relativistic oscillator has a perfect agreement in ultra-relativistic limit with observables to be obtained from the following hamiltonian function
$$
H \approx |p| + \frac{1}{2} m \omega ^ 2 q ^ 2.
$$

\section{Conclusion}
The main goal of our work is to construct relativistic
generalization of the PIMC method. This method we plan to use for
investigation of the properties of the relativistic quantum medias
with instantaneous interactions between the particles. Last fact
gives us possibility to  avoid the problem with correctness of the
many-body quantum-mechanical interaction. Of course, there are many
interesting applications one can find in Relativistic Quantum
Chemistry and Condensed Matter Physics.

To test our approach we study simple one-dimentional system
with quadratic external potential - relativistic harmonic
oscillator. This system gives us the possibility of testing our
approach because relativistic harmonic oscillator can be
studied by using Schrodinger equation in momentum space. The
comparison of the results of these two approaches has shown that our
relativistic generalization of PIMC method can be used for
investigation of quantum systems  which contain the relativistic
particles. It is very essential to emphasize that in
ultra-relativistic some special considerations are needed for
solving the problem of thermalization of configuration.

The reported study was supported by the Supercomputing Center of
Lomonosov Moscow State University \cite{parallelru}.

The work has been supported by Russian Science Foundation grant 14-22-00161.

\bibliographystyle{plain}
\bibliography{paper}

\begin{thebibliography}{0}
\bibitem{Condensed helium}
	D. M. Ceperley Path integrals in the theory of condensed helium. Rev.Mod.Phys. 67 (1995) 279-355
\bibitem{Novoselov:04:1}
  K.\,S. Novoselov, A.\,K. Geim, S.\,V. Morozov, D. Jiang, Y. Zhang,
  S.\,V. Dubonos, I.\,V. Grigorieva, and A.\,A. Firsov,
  \emph{Electric Field Effect in Atomically Thin Carbon Films},
  \emph{Science} \textbf{306} (2004) 666.
\bibitem{Fiziev}
P. P. Fiziev, \emph{Relativistic Hamiltonians wit square-roofs in
the Path Integral formaism}, \emph{Theor. Math. Phys.} {\bf 62}
(1985) 186.
\bibitem{Redmount}
I. H. Redmount and W.-M. Suen, \emph{Path integration in
relativistic quantum mechanics}, \emph{ Int. J. Mod. Phys. A} {\bf
08} (1993) 1629.
\bibitem{Filinov1}
V. S. Filinov, Yu. B. Ivanov, V. E. Fortov, M. Bonitz, and P.R.
Levashov, \emph{Color path-integral Monte-Carlo simulations of
quark-gluon plasma: Thermodynamic and transport properties},
\emph{Phys.Rev. C} {\bf 87} (2013) 035207.
\bibitem{Filinov2}
V. S. Filinov, M. Bonitz, Y. B. Ivanov, M. Ilgenfritz, and V. E.
Fortov, \emph{Thermodynamics of the quark-gluon plasma at finite
chemical potential: color path integral Monte Carlo results},
arXiv:1408.5714
\bibitem{Creutz}
	M. Creutz and B. A. Freedman, "A statistical approach to quantum mechanics," Ann. Phys. 132, 427-462 (1981).
\bibitem{Rothe}
	Heinz J Rothe Lattice Gauge Theories: An Introduction (3rd Edition) (World Scientific Lecture Notes in Physics). Published by World Scientific Publishing Co. Pte. Ltd. 2005
\bibitem{parallelru} Vl.V. Voevodin, S.A. Zhumatiy,  S.I. Sobolev, A.S. Antonov, P.A. Bryzgalov, D.A. Nikitenko,
K.S. Stefanov, Vad.V. Voevodin  \emph{Practice of "Lomonosov"
Supercomputer}, \emph{Open Systems J.} - Moscow: Open Systems
Publ., 2012, no.7.

\end{thebibliography}

\end{document}